\documentclass[preprint]{yingxiu_ma_aastex}
\usepackage{yingxiu_ma_spr-astr-addons}
\usepackage{url}

\RequirePackage{color}

\begin{document}

\title{A detailed study of the high-mass clump interacting with the bubble N10}
\shorttitle{Short article title}
\shortauthors{Yingxiu Ma et al.}

\author{Yingxiu Ma\altaffilmark{1,2}}

\author{Jianjun Zhou\altaffilmark{1,3}}
\author{Jarken. Esimbek\altaffilmark{1,3}}
\author{ Weiguang Ji\altaffilmark{1,3}}
\author{Gang Wu\altaffilmark{1,3}}
\author{Ye Yuan\altaffilmark{1,2}}

\altaffiltext{1}{Xinjiang Astronomical Observatory, Chinese Academy of Sciences, Urumqi 830011, PR China}
\altaffiltext{2}{University of Chinese Academy of Sciences,Beijing 100080, PR China}
\altaffiltext{3}{Key Laboratory of Radio Astronomy, Chinese Academy of Sciences, Urumqi 830011, PR China}

\begin{abstract}
We performed a detailed study of the high-mass clump interacting with bubble N10 based on the spectral lines $^{12}CO(3-2)$, $HCO^+(4-3)$, $N_2H^+(4-3)$ and $CH_3OH(7(0,7)-6(0,6))$ and continuum emission data at 450 $\mu$m and 850 $\mu$m released on CADC and Spitzer data. Blue-shifted optically thick line $^{12}CO (3-2)$ seems to indicate that the outer envelope of the high-mass clump is still falling toward the center. Detection of $CH_3OH(7(0,7)-6(0,6))$ suggests that
        a hot core has formed around YSO N10-7. And position-velocity diagram of $N_2H^+ (4-3)$ indicates the cold dense core of the clump has not been destroyed by the star formation activities. The mass of N10-7 is about 27.44 $M_\odot$. The ratio $HCO^+(4-3)$/$N_2H^+ (4-3)$ in the outer part of the clump is larger than that in the inner part of it. The reason may be that the CO abundance relative to $N_2H^+ (4-3)$ increased in the outer part of the high-mass clump, more $N_2H^+ (4-3)$ were converted into $HCO^+(4-3)$.
\end{abstract}


\section{Introduction}
   The expansion of the HII regions is extreme interest to studies of star formation as their expansion may trigger new generations of star formation into being within the molecular material surrounding the bubbles (Thompson et al. \cite{Thompson}). Evidence of triggering has been reported by many authors (e.g. Deharveng et al. \cite{Deharveng}; Zavagno et al. \cite{Zavagno}; Kang et al. \cite{Kang}). It should be noted, the majority of observational studies into triggered star formation near SNR or HII regions take a phenomenological approach, the evidence of triggered star formation is not very conclusive (Kendrew et al. \cite{Kendrew}). The statistical approach may address the uncertainties inherent in observations of individual HII regions. One detailed statistical study of massive star formation in the environment of 322 Spitzer mid-infrared bubbles by using the Red MSX source survey for massive young stellar objects (YSOs) suggest that the fraction of massive stars in the Milky Way formed by triggering could be between 14 and 30 per cent (Thompson et al. \cite{Thompson}). Kendrew et al. (\cite{Kendrew}) made a similar statistical study with 5106 infrared bubbles, they estimated that approximately 22 per cent of massive young stellar stars may have formed as a result of feedback from expanding HII regions.
   Therefore, the infrared dust bubbles could be good sites for us to find high-mass YSOs and study the process of high-mass star formation.

   N10 is a bright mid-infrared and radio continuum bubble with an elliptical or slight
   cometary shape with an opening at southeast of bubble (Watson et al.\cite{Watson}). Its
   kinematic distance is about 4.6 kpc (Deharveng et al.\cite{Deharveng}). The bubble is bordered on two
   sides by infrared dark clouds (IRDCs), they are interacting with the HII regions (Deharveng et al.\cite{Deharveng}).
   Especially for the IRDC in which one medium-to-high mass YSO has been found by Watson et al.\cite{Watson}, the YSO is coincident with one very dense dust core (Miettinen \cite{Miettinen}). So this IRDC is one high-mass clump and is forming YSOs. It provide us a opportunity to study initial conditions and processes of YSO's formation and triggered star formation.

   In this paper, we study the high-mass clump and high-mass YSO formed in it based on archive
   data, which including continuum data at wavelengthes 1100 $\mu$m, 850 $\mu$m, 450 $\mu$m and molecular lines of $^{12}CO(3-2)$,
   $HCO^+(4-3)$, $N_2H^+(4-3)$, $CH_3OH(7(0,7)-6(0,6))$. The observational details and data analysis are presented in the following sections.

\section{Data}
Three large-scale survey results were used in our study, which includes Galactic Legacy Infrared Mid-Plane Survey Extraordinaire (GLIMPSE), the NRAO VLA Sky Survey (NVSS) and BOLOCAM Galactic Plane Survey (BGPS). GLIMPSE is performed using the Spitzer Space Telescope by Spitzer-Infrared Array Camera (IRAC). It is a mid-infrared survey of the inner Galaxy at 3.6, 4.5, 5.8 and 8 $\mu$m with angular resolution of between 1\farcs5 and 1\farcs9, and a very large catalog of stars were extracted from the survey (Churchwell et al.\cite{Churchwell}). Here we used the mosaicked images from GIMPSE acquired by Spitzer-IRAC at 8 $\mu$m and the Point-Source catalog. NVSS is a 1.4 GHz radio continuum survey covering the entire sky north of -40 deg declination with an angular resolution of about 45\farcs (Condon et al.\cite{Condon}). BGPS is a 1.1 mm continuum survey of the Galactic Plane made using Bolocam on the Caltech submillimeter Observatory, the survey coverage totals 170 square degrees (with 33\farcs FWHM effective resolution) and detected about 8400 point sources (Rosolowsky et al \cite{Rosolowsky}).

   In addition, we used the James Clerk Maxwell Telescope (JCMT) data released on  Canadian Astronomy Data Centre
 (CADC), they
   including $^{12}CO(3-2)$, $HCO^+(4-3)$, $N_2H^+(4-3)$, $CH_3OH(7(0,7)-6(0,6))$ data and dust continuum
    emission at 450 $\mu$m and 850 $\mu$m. The co-added spectral
     cubes were binned to 0.2 $km\, s^{-1}$, and the data were smoothed with
     6\farcs Gaussian kernel. The final spatial resolution of each cube is
      16\farcs (Smith et al.\cite{Smith}). The angular resolution of 450 $\mu$m
      and 850 $\mu$m is 8\farcs5 and 14\farcs5 respectively (Leeuw\& Robson
      \cite{Leeuw}).
\section{Results and discussion}
\subsection{N10}

\begin{figure}[htbp]
 \includegraphics[width=5.5cm,angle=90]{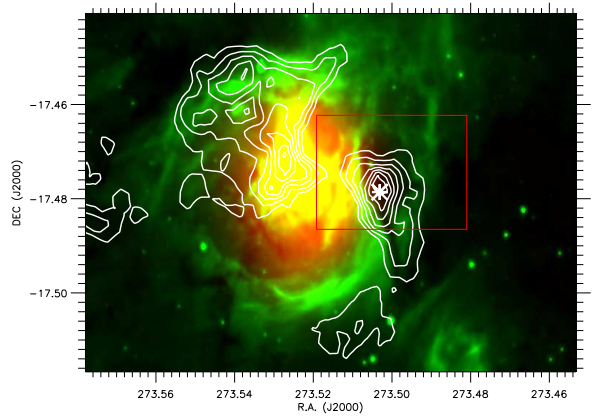}
 \caption{ N10, the $^{12}CO$ (3-2) contour levels range from 90.0 to 159.5 by 9.93 $K \,km \,s^{-1}$ are superimposed on the composite colour image with 20 cm emission in red and 8 $\mu$m emission in green. The high-mass clump which we study in detail was marked with one red square. The white star denotes the YSO (N10-7) identified by Watson et al. \cite{Watson}.}
\label{figa}
\end{figure}
Figure 1 shows the $^{12}CO(3-2)$ emission contours over the two colour image
    with 20 cm emission in red and 8 $\mu$m emission in green. The border of N10
    is well traced by the 8 $\mu$m emission. The 20cm emission in the bubble indicates that there is a large HII region ionized
    by central stars. The high-mass clump, which we will study in detail, was marked with one red square. An obvious arch border appears between the bubble and the high-mass clump, which suggests that there exists strong interaction. One very compact CO cloud is coincident with the high-mass clump, the medium-to-high-mass YSO (N10-7) identified by Watson et al. \cite{Watson} is located at the center of it.
     The more extended CO cloud
     appear on the upper left to the center of N10. The CO emission toward the center of the bubble in projection could be arising from the front or back side of the bubble.

    We estimated the dynamical age of N10 using the model described by Dyson \&
    Williams \cite{Dyson} with a given radius R as:
    \begin{eqnarray}
     t(R) =\frac{4\,R_s}{7\,c_s}\left[(\frac{R}{R_s})^{\frac{7}{4}}-1\right]
    \end{eqnarray}

     where $c_s$ is the sound velocity in the ionized gas ($c_s = 10 km\, s^{-1}$) and
     $R_s$ is the radius of the Str\"{o}mgren sphere given by $R_s = (3N_{Lyc}/4\pi\,
     n_0^2\,\alpha_\beta)^{\frac{1}{3}}$, where $N_{Lyc}$ is the number of ionizing
     photons emitted by the star per second, $n_0$ is the original ambient density,
     and $\alpha_\beta = 2.6 \times 10^{-13} cm^3\, s^{-1}$ is the hydrogen
     recombination coefficient to all levels above the ground level. Here we adopt
      a Lyman continuum photon flux of about $2.2 \times 10^{49} ph\, s^{-1}$
      (Watson et al.\cite{Watson}) and a radius of 1.61 pc. For that no CO cores look like swept by expanding HII region on the border of N10, it is difficult for us to estimate the original ambient density of N10. Massive star formations usually take place in the giant molecular clouds and the densities of the giant molecular clouds are about $10^3$ to $10^4 cm^{-3}$ (Goldsmith.\cite{Goldsmith}). So it seems to be safe for us to assume the low value of original ambient density of N10 is about $10^3 cm^{-3}$ and obtain a lower bound on the dynamical age,about $9.17 \times 10^4$ yr. Taking into consider that stars are formed in the dense clumps of the giant molecular cloud, the true original ambient density may be larger than that we assumed, the true dynamical age of N10 should be larger than $9.17 \times 10^4$ yr.
\subsection{The High-mass Clump}

\begin{figure}[htbp]
\includegraphics[width=6cm,angle=90,scale=0.9]{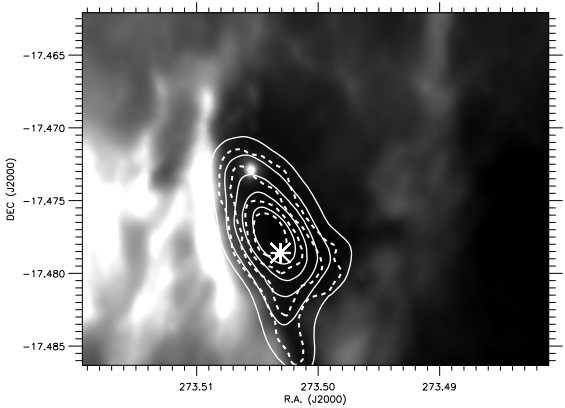}
\caption{ The white solid contours denote 850 $\mu$m emission, contour levels range from 0.77 to 3.43 by 0.53 Jy / beam. The white dotted contours denote 450 $\mu$m, contour levels range from 3.57 to 11.95 by 2.10 Jy / beam. The lowest levels of 850 $\mu$m and 450 $\mu$m are both larger than $3\sigma$. The background is 8 $\mu$m emission. The star represents N10-7.}
\label{figb}
\end{figure}
\begin{figure}[htbp]
\includegraphics[width=6cm,angle=90,scale=0.9]{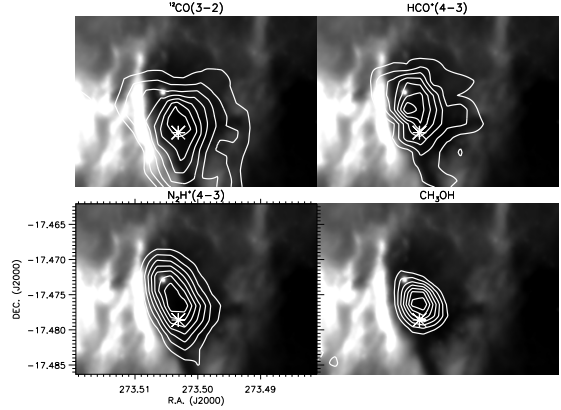}
\caption{$^{12}CO(3-2)$, $HCO^+(4-3)$, $N_2H^+(4-3)$ and $CH_3OH(7(0,7)-6(0,6))$ contours map of the clump are superimposed on the 8 $\mu$m emission. The contour levels range from 90.0 to 159.5 by 9.93 $K \,km \,s^{-1}$ for $^{12}CO(3-2)$, 3.0 to 15.78 by 1.83 $K \,km \,s^{-1}$ for $HCO^+$, 3.0 to 14.43 by 1.63 $K \,km \,s^{-1}$ for $N_2H^+$ and 3.0 to 17.15 by 2.02 $K \,km \,s^{-1}$ for $CH_3OH(7(0,7)-6(0,6))$. The star represents N10-7.}

\label{figc}
\end{figure}
Past continuum observations at differen wavelength have proven that there is one dust condensation in the
   high-mass clump, the YSO (N10-7) is forming in it (Miettinen \cite{Miettinen}).
   The JCMT continuum observation results at 450 $\mu$m
   (white dotted contours) and 850 $\mu$m (white solid contours)
   display a elongated dust core from north to south (see Fig.2).
   N10-7 is nearly at the center of it.
   No mid-IR diffuse emission and
    lack of 20 cm continuum emission suggest that high-mass star formation is in its very
    early stage, or the high-mass YSO was embedded in the clump very deeply.

    The total dust mass of the clump is
    calculated using $M = S_vD^2/k_v B_v(T_d)$, where $S_v$ is the flux at
    frequency $v$, D is the distance (4.6 kpc), $B_v(T_d)$ is the plank function,
    and $k_v$ is the dust opacity per unit gas /dust mass. $T_d$ is the dust temperature, it was obtained by SED fitting with a simple greybody model.  The total flux of the clump at 450 $\mu$m, 850 $\mu$m and 1.1 mm are about 99.33, 13.30 and 7.6 Jy, respectively. The flux at 450 and 850 $\mu$m were derived from JCMT data by ourself, the flux at 1.1 mm were obtained from BGPS point source catalog and multiplied by a factor of 1.5. The best fitted result indicates that $T_d = 27$ K. Using the dust opacity
    of $k_{850} = 0.02 cm^2 g^{-1}$ obtained from Ossenkopf \& Henning\cite{Ossenkopf}, the ratio of dust to gas is taken as 0.01. We derived a total mass
      of $M = 908 M_\odot$ from 850$\mu$m flux. The area of the clump is 1333 arcsec square. Assuming the clump is spherical, we get
       the core radius of 20\farcs6 (corresponding to 0.46 pc). Then the average number density is
          $4.59 \times 10^4cm^{-3}$. Following Beuther et al.\cite{Beuther12} and Longmore et al.\cite{Longmore}, we assume $\rho$$\propto$$r^{-2}$ and calculate Virial mass using the formula of Maclaren, Richardson and Wolfendale\cite{Maclaren}.
   \begin{eqnarray}
    M_{vir} =126\,R\,\Delta\,V^{2}
   \end{eqnarray}
   where $\bigtriangleup$V is the full width at half-maximum intensity in $km s^{-1}$, R is the radius of the molecular cloud in parsecs, M is cloud mass in unit of solar mass. Since star is formed in the cold dense core traced by $N_2H^{+}$, we used the Gaussian line width of $N_2H^{+}$ obtained at the position of YSO (N10-7) to calculate the Virial mass of the
   clump. Finally, we obtained $M_{vir}$ =$1640 M_\odot$. This indicate the clump may be gravitational stable. However, Miettinen \cite{Miettinen} obtained a total mass of $3335\underline{+}704 M_\odot$ for the clump based on 870 $\mu$m observation with Apex submillimeter telescope. The mass difference may be due to the observation sensitivity difference between the two telescopes. If the result of Miettinen \cite{Miettinen} is true, the clump tends be under gravitational unstable. Taking into that the clump is forming one YSO in it, it seems to be more reasonable. Just as that pointed out by Beuther et al.\cite{Beuther12}, there is large uncertainties for the mass estimated from the dust continuum emission and Virial mass calculated with $N_2H^{+}$, such a result are not very conclusive.

 \subsection{The kinematics of the clump}

    \begin{figure}[htbp]
    \includegraphics[width=5.6cm,angle=90]{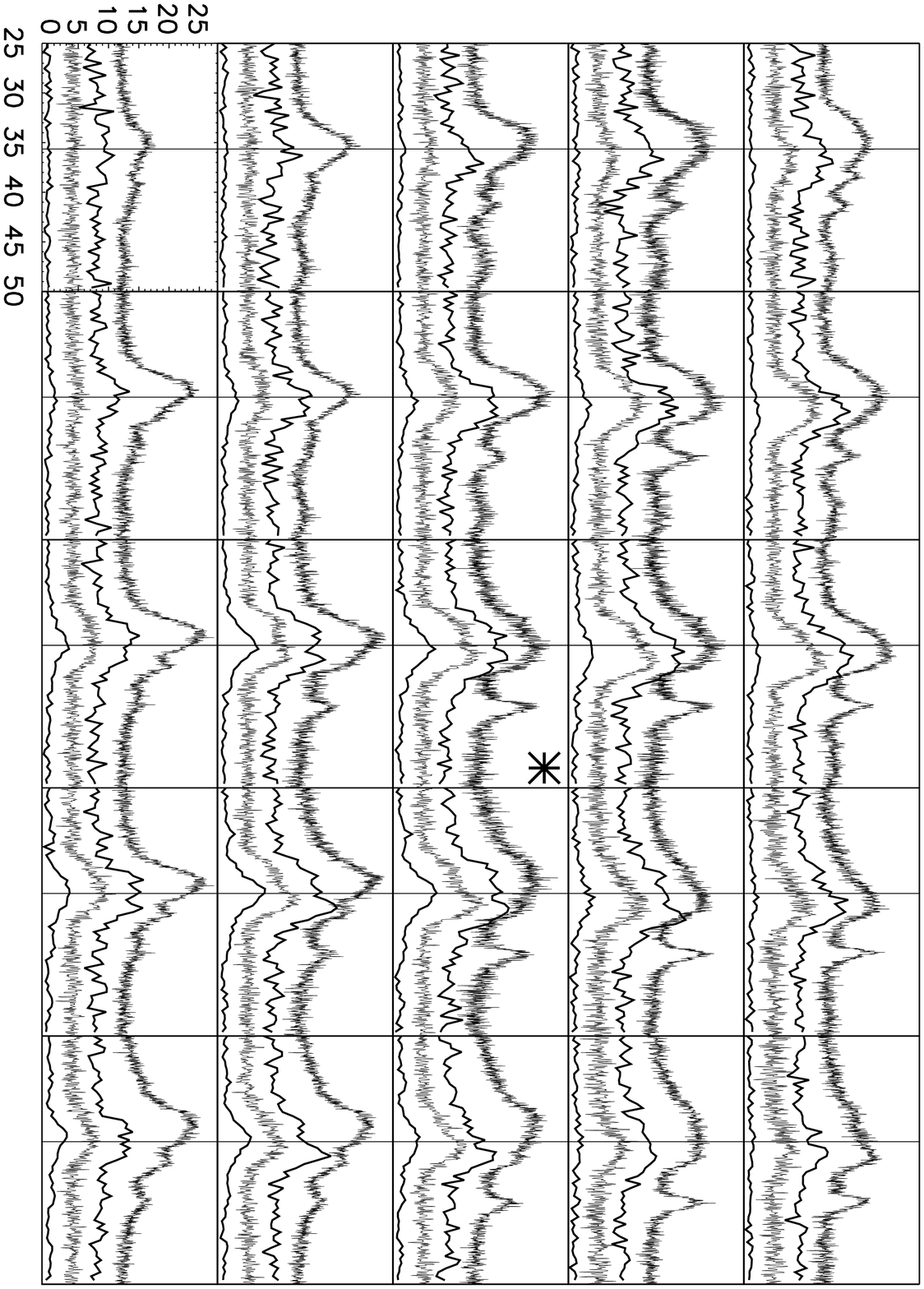}

    \caption{The molecular profiles extracted from the positions around the YSO N10-7 are plotted, the panel marked by a star correspond to the YSO N10-7. From up to down in each panel is $^{12}CO (3-2)$, $HCO^+ (4-3)$, $N_2H^+ (4-3)$, $CH_3OH(7(0,7)-6(0,6))$, respectively. The velocity is between 25 to 50 km/s. $CH_3OH$, $N_2H^+$, $HCO^+$ profiles are scaled up by a factor 5, 3 and 5, respectively, and their base line is added 4, 8, 12 $K\, km\, s^{-1}$, respectively. The vertical line indicates the system velocity of the high-mass clump which is 35.64 $km\, s^{-1}$.}

    \label{figd}
    \end{figure}
     $N_2H^+$ is
   not strongly affected by freeze out on grain surfaces, it is a good tracer of younger cold and massive
   cloud cores. $HCO^+$ is a
   highly abundant molecule, with abundance especially enhanced around regions of
   higher fractional ionization. it is also enhanced by the presence of outflows
   where shock-generated radiation fields are present (Vasyunina et al.\cite{Vasyunina}).
   $CH_3OH$ molecules are formed in the icy
   mantles of interstellar dust grains by hydrogenation of CO molecules at temperature
   of 10K, as the young protostellar object evolve and warm up its environment the
   $CH_3OH$ molecule sublimate from the dust grains into the gas phase at 100K (Torstensson\cite{Torstensson}).
  Detection of $HCO^+(4-3)$ and $CH_3OH$ toward the high-mass clump
   indicates that one hot core has formed around the YSO N10-7. On the other hand, detection of $N_2H^+(4-3)$ indicates star formation in the clump is
   still in its early stage, the natal molecular cloud was not destroyed by the feedback of YSO (see Fig.3).

   Fig.4 displays the $^{12}CO (3-2)$, $HCO^+ (4-3)$, $N_2H^+ (4-3)$, $CH_3OH(7(0,7)-6(0,6))$
     profiles according to their position around the YSO
     N10-7. Here $CH_3OH(7(0,7)-6(0,6))$, $N_2H^+(4-3)$ and $HCO^+(4-3)$ profiles are scaled up by a factor of 5, 3 and 5, respectively. The panel marked by a star indicates the position of YSO N10-7.
     the vertical line in the map indicates the system velocity (35.64 $km\, s^{-1}$) of the high-mass clump.
     We noted that the profiles of optically thick line $^{12}CO (3-2)$ include two components at velocity 34.57 and 41.77 $km\, s^{-1}$. The first is the stronger one and is usually blue-shifted
     with respect to the system velocity. This component traces the high-mass clump, the blue-shifted profiles of it may indicate that the outer envelope traced by CO is still falling toward the center. The profiles of $HCO^+(4-3)$ in the all upper two rows and the left two panels of the third and fourth rows are red-shifted with respect to the system velocity. The $N_2H^+ (4-3)$ profile at the position of YSO show evidence of self-absorption dip. The profiles in upper two rows of panels
     are red-shifted with respect to the system velocity, while in the lower three rows of panels they become blue-shifted. The position-velocity diagram of $N_2H^+(4-3)$ shows a velocity gradient along the major axis
     of the high-mass clump (see Fig.5). Such a velocity gradient may
     be a combination of both rotation and infall motions of the cold dense core (Tobin et al.\cite{Tobin}). This supports our idea that cold dense core of the clump has not been destroyed by the feedback of the YSO N10-7 yet.
     The optically thin $CH_3OH(7(0,7)-6(0,6))$ profiles have only one peak with obvious line wings, but corresponding
     position-velocity diagram (see
     Fig.6) display only slightly red-shifted and blue-shifted lobes. In some extent, this supports the exist of outflow. However, other motions such as rotation, infall and turbulence could also cause this effect,we can not exclude these motions. More high-resolution observations and studies are needed to draw a conclusive result.

     \begin{figure}[htbp]
     \includegraphics[width=6cm]{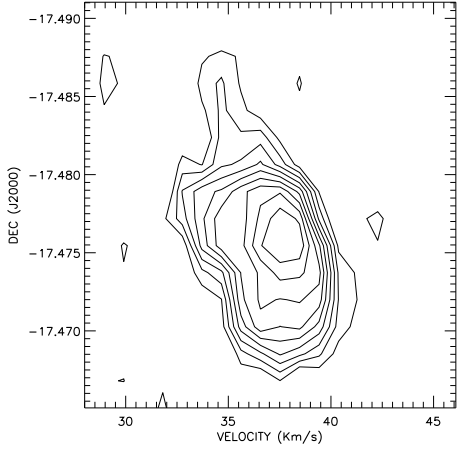}

     \caption{ PV diagram of $N_2H^+ (4-3)$ emission along the major axis of the high-mass clump. The contours are from 10\% to 90\% with steps of 10\% of the peak of the integrated intensity.}

     \label{fige}
     \end{figure}

     \begin{figure}[htbp]
     \includegraphics[width=6.0cm]{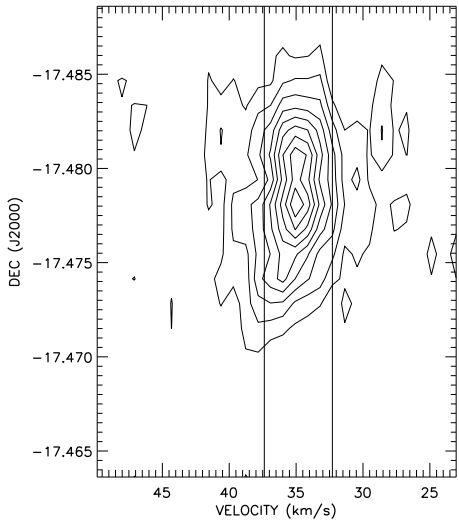}

     \caption{PV diagram for the $CH_3OH(7(0,7)-6(0,6))$ emission along the major axis of the high-mass clump. The contours are from 10\% to 90\% with steps of 10\% of the peak of the integrated intensity. The vertical black lines indicate the blue- and red-shifted components' velocity boundary. }

     \label{figf}
     \end{figure}
       Assuming that line wings of $CH_3OH(7(0,7)-6(0,6))$ indicate outflows truly, we plot the blue and red lobe of the outflow (see Fig.7),
        the distribution of red lobe is much larger than that of blue
        lobe. The background is the ratio $HCO^+(4-3)$/$N_2H^+ (4-3)$.
        It is interesting to find that ratio $HCO^+(4-3)$/$N_2H^+ (4-3)$ in the outer part is larger than that in the inner part of the clump. For that $HCO^+$ is created by the reaction between $N_2H^+$
        and CO. The reason may be that the CO abundance relative to $N_2H^+ (4-3)$ increased in the outer part of the high-mass clump, more $N_2H^+ (4-3)$ were converted into $HCO^+(4-3)$. Increasing of the relative abundance of $HCO^+(4-3)$ increases the ratio of $HCO^+(4-3)$/$N_2H^+ (4-3)$. The distribution of $HCO^+(4-3)$ is nearly the same as that of red lobe. This is consistent with the conclusion that $HCO^+(4-3)$ abundance will be enhanced by the presence of outflows. No 20cm continuum emission was detected to the clump suggests that it could not be highly fractional ionized. Enhanced HCO+(4-3) abundance is likely caused by the outflow. This is also consistent with the fact that most profiles of $HCO^+(4-3)$ are red-shifted with respect to the system velocity.

     \begin{figure}[htbp]
     \includegraphics[width=6cm,scale=0.8]{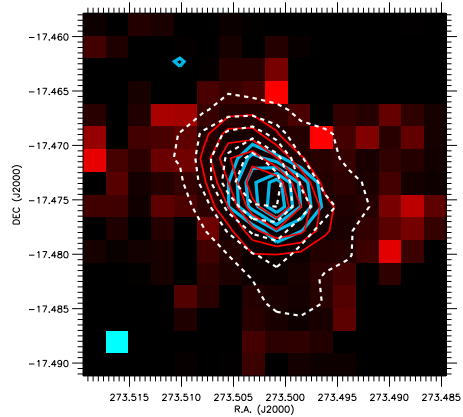}

     \caption{The components of the $CH_3OH$ outflow superimposed on the integrated intensity ratio $HCO^+(4-3)$/$N_2H^+ (4-3)$ (red background). The velocity components is integrated in (27,32.5) and (37.4,42.5) $ km\, s^{-1}$ for its blue and redshifted lobes, the blue and red contours represent blue lobe and red lobe, respectively. The white contours denote $HCO^+(4-3)$ molecule's emission.}

     \label{figg}
     \end{figure}

\subsection{SED fitting of the central massive star}

 Watson et al.\cite{Watson} fitted the SED of the young stellar object (N10-7) formed in the high-mass
 clump using the tool developed by Robitaille et al.\cite{Robitaille06} based on the Spitzer IRAC data,
 they found that N10-7 is one Stage I YSO
 with a mass of 12.4 $M_\odot$.
 However, their SED fitting results may not be very accurate due to the lack of data at longer wavelengthes.
 We fit the SED of N10-7 using more data, which including
 Spitzer-IRAC data at 3.6, 4.5, 5.8 and 8 $\mu$m, SCUBA data at 450 and 850 $\mu$m and BOLOCAM data at 1.1 mm.
 The visual extinction is estimated from
   the classical relations
   $N_{H+H_2}/E(B-V) = 5.8 \times 10^{21} particles cm^{-2} mag^{-1}$
   (Bohlin et al.\cite{Bohlin}) and $A_v = 3.1 E(B-V)$, so we
   obtain $A_v = 5.34 \times 10^{-22} N_{H_2}$. According to Shetty et
   al. \cite{shetty} and Ji et al.\cite{Ji}, we used the X factor between $H_2$ and $^{12}CO$ to estimate
    the column density of the clump. So $N_{H_2} = X \times W_{CO} [cm^{-2}]$,
    where $W_{CO}$ is the integrated $^{12}CO$ intensity, $W_{CO} = \int T_{mb}\, dv\, cm^{-2}\, K\, km\, s^{-1}$.
    Finally we obtained the $A_v$ is between 5 to 30 mag. We selected the best SED
    fitting model (see Fig.8).
    The source age is $1.81 \times 10^4 yr$, the source mass is $M_\star = 27.44 M_\odot$, the disk
    mass $M_{disk} = 4.52 \times 10^{-2} M_\odot$, the envelope mass $M_{env} =1.48 \times 10^3 M_\odot$,
    and the envelope accretion rate $\dot{M}_{env} = 3.34 \times 10^{-3} M_\odot/yr$.
    N10-7 satisfy the criteria $\dot{M}_{env}/M_\star > 10^{-6} yr^{-1}$ (Robitaille et al.(\cite{Robitaille07}), and is one stage I
     YSO.

\begin{figure}[h]
    \includegraphics[width=7.0cm]{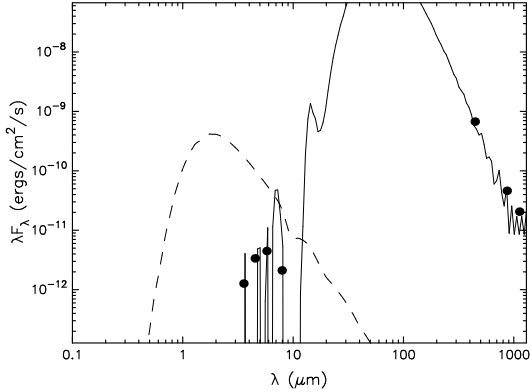}

    \caption{ The SED fitting of high-mass YSO N10-7. Black line shows the best fit. The dashed line shows the stellar photosphere corresponding to the central source of the best fitting model, as it would look in the absence of circumstellar dust. The black points represent the input data at 3.6, 4.5, 5.8, 8, 450, 850 and 1100 $\mu$m. }

    \label{figh}
    \end{figure}

 \section{Summary}

We studied the high-mass clump associated with the bubble N10 in
detail based on the Spitzer data, continuum observations at 450, 850
and 1100 $\mu$m and spectral lines of $^{12}CO (3-2)$, $HCO^+ (4-3)$,
$N_2H^+ (4-3)$ and $CH_3OH(7(0,7)-6(0,6))$.

We draw some tentative results for the high-mass clump. Optically thick line $^{12}CO (3-2)$ are all blue-shifted with respect to
        the system velocity, this seems to indicate that the outer envelope of the high-mass clump is still falling toward the center.
        The position-velocity diagram of $N_2H^+ (4-3)$ indicates the cold dense core of the clump has not been destroyed by the star formation activities in it. Detection of $CH_3OH(7(0,7)-6(0,6))$ suggests that
        a hot core has formed due to the feedback of YSO N10-7. The mass of N10-7 is about 27.44 $M_\odot$. The line wings of $CH_3OH(7(0,7)-6(0,6))$ seems to trace the outflow excited by N10-7. The ratio $HCO^+(4-3)$/$N_2H^+ (4-3)$ in the outer part of the clump is larger than that in the inner part of it. The reason may be that the CO abundance relative to $N_2H^+ (4-3)$ increased in the outer part of the high-mass clump, more $N_2H^+ (4-3)$ were converted into $HCO^+(4-3)$.

\acknowledgments
This work was funded by The National Natural Science Foundation of China under grant 10778703, and partly supported by China Ministry of Science and Technology under State Key Development Program for Basic Research (2012CB821800) and The National Natural Science Foundation of China under grant 10873025.



\end{document}